\documentclass[journal]{IEEEtran}
\usepackage{caption}
\usepackage{amsmath,amsfonts}
\usepackage{graphicx}
\usepackage{cite}
\usepackage{booktabs} 
\usepackage{url}
\usepackage[T1]{fontenc} 
\usepackage{lmodern} 
\usepackage{subcaption} 

\begin{document}

\title{Profiling Multi-Level Operator Costs for Bottleneck Diagnosis in High-Speed Data Planes}

\author{Zhiyuan~Ren,~\IEEEmembership{Member,~IEEE},
        Yutao~Liu,
        Wenchi~Cheng,~\IEEEmembership{Senior~Member,~IEEE},
        and~Kun~Yang,~\IEEEmembership{Fellow,~IEEE}
\thanks{This work was supported by the National Key Research and Development Program of China (No. 2024YFE0200300).}%
\thanks{Zhiyuan Ren, Yutao Liu, and Wenchi Cheng are with the School of Telecommunications Engineering, Xidian University, Xi'an 710071, China.}
\thanks{Kun Yang is with the School of Computer Science and Electronic Engineering, University of Essex, Colchester CO4 3SQ, U.K.}
\thanks{Corresponding author: Zhiyuan Ren (zyren@xidian.edu.cn).}%
}

\markboth{IEEE Networking Letters,~Vol.~XX, No.~X, Month~2025}%
{Author \MakeLowercase{\textit{et al.}}: }

\maketitle

\begin{abstract}
This paper proposes a saturation throughput delta–based methodology to precisely measure operator costs in high-speed data planes without intrusive instrumentation. The approach captures non-linear scaling, revealing that compute-intensive operators like CRC exhibit super-linear behavior, while most others are sub-linear. We introduce the Operator Performance Quadrant (OPQ) framework to classify operators by base and scaling costs, exposing a cross-architecture Quadrant Shift between Arm and x86. This method provides accurate, architecture-aware bottleneck diagnosis and a realistic basis for performance modeling and optimization.
\end{abstract}


\section{Introduction}
\IEEEPARstart{W}{ith} the rapid development of 5G and cloud computing, software-defined high-speed data planes (e.g., DPDK, VPP) have become the core engines of modern network infrastructure. A precise, quantitative understanding of each network function operator's performance cost is a prerequisite for identifying system bottlenecks and guiding optimizations. However, accurately measuring these fine-grained operators on complex modern processors presents significant challenges.

These challenges expose the limitations of existing approaches. Intrusive techniques risk altering performance, while passive monitoring and system-level tools often focus on different objectives or lack the required granularity. For instance, some non-intrusive methods excel at measuring per-flow latency by using sophisticated statistical techniques \cite{shahzad_accurate_2016}, and practical tools like FloWatcher-DPDK provide lightweight flow-level monitoring \cite{zhang_flowatcher-dpdk_2019}, but these do not directly reveal the computational cost of individual operators. Foundational performance modeling work has established that the throughput of high-performance I/O frameworks like DPDK is fundamentally limited by CPU cycles, providing a model to measure a framework's intrinsic overhead \cite{gallenmuller_comparison_2015}. This modeling approach has been extended to more complex, heterogeneous hardware like SmartNICs \cite{guo_lognic_2023}. While these models offer powerful predictive capabilities, they often assume linear costs or require complex, white-box parameterization. Furthermore, the challenge of performance portability is well-recognized; benchmarking studies consistently show that performance profiles of virtualization technologies differ significantly across architectures like x86 and Arm \cite{acharya_performance_2018}, and radical new hardware such as memory networks has been proposed to overcome fundamental memory bottlenecks \cite{korikawa_memory_nodate}. This need for precise performance understanding is underscored by its application in demanding real-time control systems \cite{anastasio_integration_2025} and its direct impact on the techno-economic TCO models that guide network deployment \cite{bouras_cost_2016, bista_vnf_2022}.

To overcome these limitations, this paper proposes a methodology based on saturation throughput delta that is externally observed, involving no in-pipeline instrumentation, which accurately deduces an operator's true cycle cost without interfering with its execution environment. Our analysis reveals a more complex reality than prior models assume: compute-intensive operators exhibit a distinct super-linear (accelerating) scaling characteristic, while most other operators follow a sub-linear (decelerating) trend. To analyze this, we developed the OPQ framework. Based on this, we profiled operators on x86 and Arm platforms and observed a significant Quadrant Shift phenomenon, providing critical insights for cross-platform optimization. 

The main contributions of this study are:
(1) a non-intrusive, high-precision methodology for diagnosing operator costs based on saturation throughput delta;
(2) identification of a clear scaling dichotomy, with compute-intensive operators (e.g., CRC) showing super-linear growth and most others sub-linear;
(3) the Operator Performance Quadrant (OPQ) framework for classifying operators by base and scaling costs; and
(4) data-driven guidelines for addressing critical bottlenecks and enabling cross-platform optimization.


\section{Measurement Methodology}

\subsection{Saturation-Based Measurement Protocol}
To achieve precise and perturbation-free measurement, we designed a protocol based on the assumption that a saturated data plane is limited by the CPU's clock cycle resources. This creates a deterministic relationship between external throughput ($R$) and internal per-packet cost ($C$). In this context, throughput R is measured in packets per second (pps).

The protocol involves two core steps:
\begin{enumerate}
    \item \textbf{Baseline Measurement}: We measure the maximum saturation throughput ($R_{base}$) of a minimal L2 forwarding system. Its cost ($C_{base}$) satisfies:
    \begin{equation}
        R_{base} \times C_{base} = F_{cpu} \label{eq:base}
    \end{equation}

    \item \textbf{System Under Test (SUT) Measurement}: We insert a single target operator, \textbf{op}, and measure the new saturation throughput, $R_{op}$. The new cost is:
    \begin{equation}
        R_{op} \times (C_{base} + C_{op}) = F_{cpu} \label{eq:sut}
    \end{equation}
\end{enumerate}

By solving (\ref{eq:base}) and (\ref{eq:sut}), we derive the operators' cost, $C_{op}$:
\begin{equation}
    C_{op} = F_{cpu} \times \left(\frac{1}{R_{op}} - \frac{1}{R_{base}}\right) \label{eq:final}
\end{equation}
This protocol is repeated for various packet sizes to study cost scaling.

\subsection{Methodological Rigor and Environmental Controls}

To ensure that CPU cycles are the sole systemic bottleneck—a core assumption of our methodology—we employ three key environmental controls: binding the application to dedicated cores, isolating network IRQs to non-experimental cores, and disabling dynamic CPU frequency scaling to maintain a constant $F_{cpu}$.

Our methodology assumes that the cost of the base system ($C_{base}$) remains constant when a new operator is introduced. While adding an operator can introduce minor perturbations to the baseline cost $C_{base}$, such as instruction cache contention, our repeated measurements confirmed this impact to be negligible.

\subsection{OPQ Analysis Framework}
To translate raw cost data ($C_{op}$) into actionable insights, we constructed the OPQ framework based on two dimensions:

\begin{enumerate}
    \item \textbf{Base Cost (Y-axis)}: The operator's fixed overhead, defined as its net cost for the smallest packet size, $C_{op}(s=64)$.
    \item \textbf{Scaling Behavior (X-axis)}: The trend of cost growth is characterized by fitting the net operator cost data to a power-law model. The exponent $k$ from this fit serves as a robust indicator of the scaling behavior, defined as:
    \begin{equation}
        C_{op}(s) = a \cdot s^k \label{eq:power_law}
    \end{equation}
    where $s$ is the packet size. The value of $k$ directly classifies the operator's scaling model: $k>1$ indicates accelerating (super-linear) cost, while $k<1$ indicates decelerating (sub-linear) cost.
\end{enumerate}

Based on these dimensions, we classify operators into four quadrants. The horizontal axis is divided at k = 1 (accelerating vs. decelerating), while the vertical axis (Base Cost) is split by the dataset median. While operators near a boundary might shift with a different dataset, our key findings, such as the major shifts of CRC and hash, are robust and well beyond minor boundary perturbations. The quadrants guide optimization strategies, from batching for High Startup Cost operators to algorithm replacement for Latent Traps.

We acknowledge certain methodological nuances: the CPU-as-sole-bottleneck model is an approximation for I/O-bound or extremely low-cost operators, and the observed Quadrant Shift is a mixed effect of multiple architectural and compiler factors beyond clock speed.

\section{Cross-Architecture Performance Analysis}

\subsection{Experimental Setup}
To facilitate a direct cross-architecture comparison, we deployed our measurement protocol on two distinct hardware platforms:
\begin{itemize}
    \item \textbf{Arm Platform:} A Marvell CN96XX  server (1.8 GHz).
    \item \textbf{x86 Platform:} Intel Xeon Silver 4210 (2.20 GHz) .
\end{itemize}
Both systems were equipped with a 100GbE NIC and ran Ubuntu 22.04 with DPDK 21.11. To ensure that the CPU was the sole bottleneck as required by our methodology, all throughput measurements were conducted using an external 100Gbps traffic generator and analyzer. This high-speed instrumentation guarantees that the measurement process itself does not become a limiting factor, allowing the system's true CPU saturation points ($R_{base}$ and $R_{op}$) to be accurately determined. The tested operators, chosen for their prevalence in network functions, are described and classified in Table~\ref{tab:operator_class}. The full measurement protocol was executed for 64, 128, and 256-byte UDP packets on both platforms under identical environmental controls as described in Section~II-B.

\begin{table}[htbp]
\centering
\caption{Classification and Description of Profiled Operators}
\label{tab:operator_class}
\begin{tabular}{llp{3cm}}
\toprule
\textbf{Operator} & \textbf{Class} & \textbf{Functional Description} \\
\midrule
\texttt{CRC}      & Compute-intensive         & Cyclic Redundancy Check calculation for frame integrity. \\
\texttt{hash}     & Memory-intensive          & Hash table key lookup/insertion for flow identification. \\
\texttt{htons}    & Simple data manipulation  & Converts a 16-bit value from host to network byte order. \\
\texttt{Checksum} & Compute-intensive         & Calculates the Internet Checksum for header validation. \\
\texttt{printf}   & System I/O call           & Formatted output via a kernel context-switching system call. \\
\texttt{rte\_log}  & Optimized logging         & DPDK's lightweight logging function to avoid kernel overhead. \\
\bottomrule
\end{tabular}
\end{table}

\subsection{Performance Profiles and Key Observations}

\begin{figure}[hbtp]
\centering
\begin{subfigure}[b]{0.48\textwidth}
    \includegraphics[width=\textwidth]{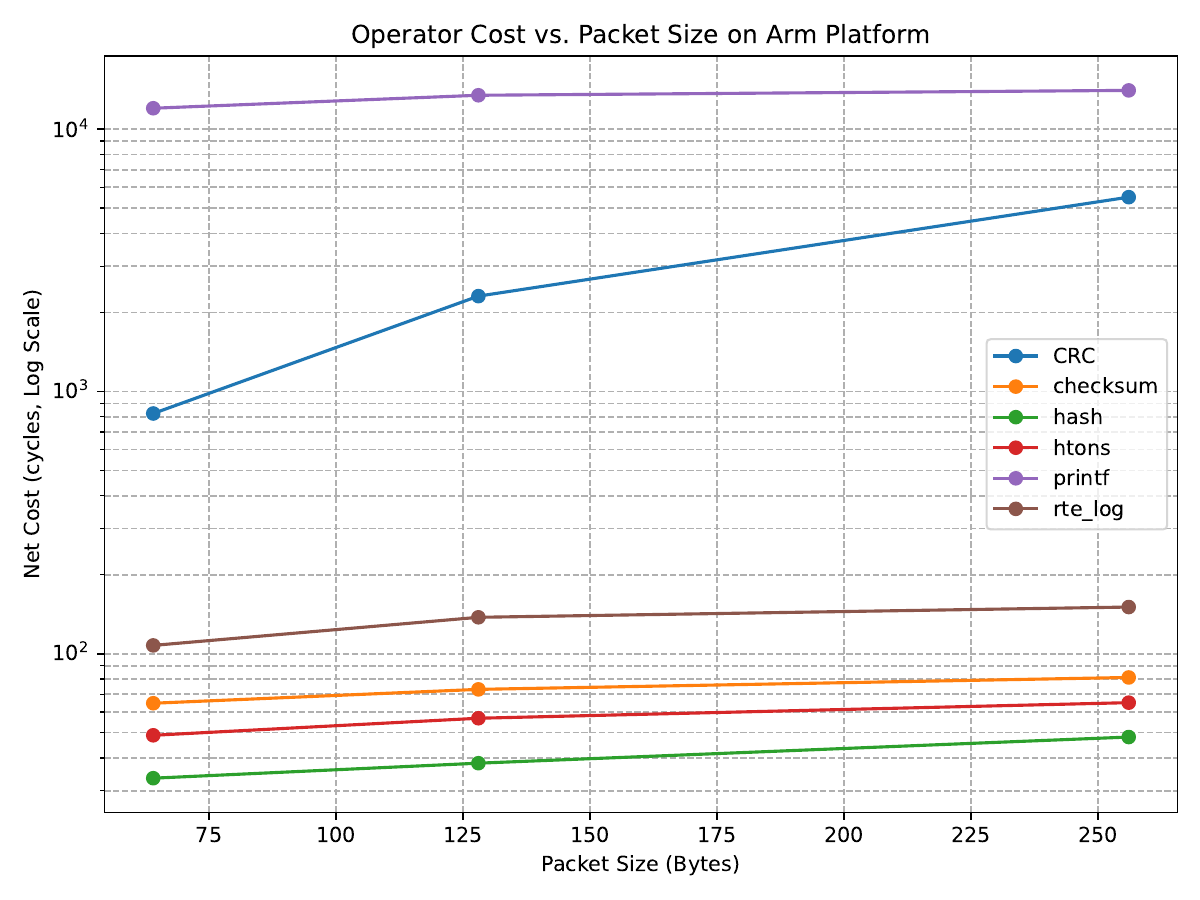}
    \caption{Arm Platform}
    \label{fig:cost_scaling_arm}
\end{subfigure}
\hfill 
\begin{subfigure}[b]{0.48\textwidth}
    \includegraphics[width=\textwidth]{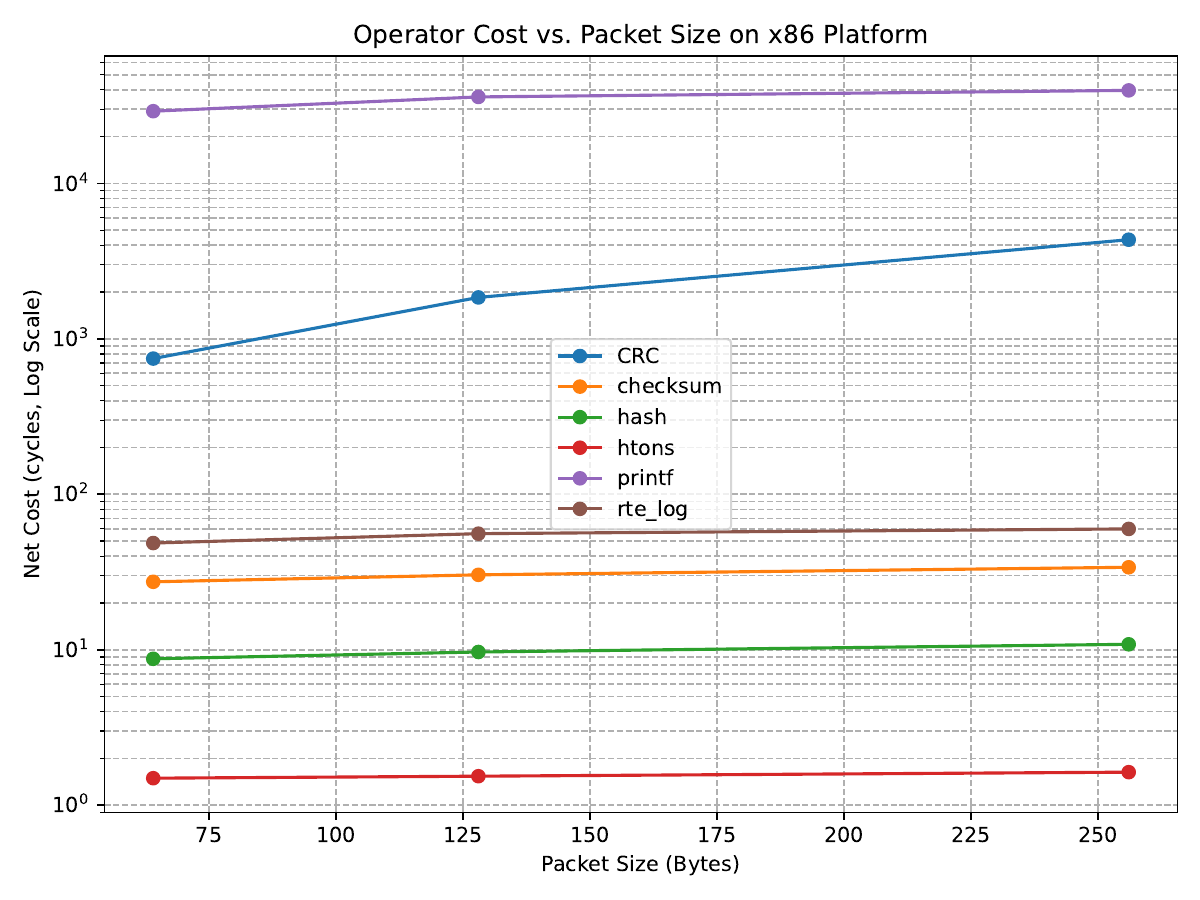}
    \caption{x86 Platform}
    \label{fig:cost_scaling_x86}
\end{subfigure}
\caption{Non-linear scaling characteristic of operator cost on both platforms. The Y-axis uses a logarithmic scale.}
\label{fig:cost_scaling}
\end{figure}

Our analysis reveals that operator scaling behavior is not universally super-linear but rather class-dependent. By fitting the net cost data to a power-law model ($Cost\propto s^k$), we identify two distinct trends detailed in Table~\ref{tab:combined_profile}: a single compute-intensive operator, \texttt{CRC}, exhibits a super-linear (accelerating) characteristic where the scaling exponent $k>1$. In contrast, the majority of other operators show a sub-linear (decelerating) trend ($k<1$), indicating that their per-packet cost increase diminishes as packet size grows, as shown in Fig. \ref{fig:cost_scaling}.

The super-linear behavior of \texttt{CRC} is likely caused by CPU cache hierarchy effects, where processing larger packets increases L1/L2 cache misses, forcing more frequent accesses to slower memory and thus raising the average cost per byte. This accelerating cost model identifies \texttt{CRC} as a primary bottleneck for workloads involving large packets.

We also identified an operator that acts as a severe performance sink. As shown in Fig. 2, the \texttt{printf} system call exhibits an extreme startup cost due to I/O and kernel context-switching overhead. Its recommended DPDK alternative, \texttt{rte\_log}, reduces this cost by over 110-fold on Arm and approximately 595-fold on x86, highlighting a critical optimization for production data planes.

It is also noteworthy that these performance characteristics differ significantly across architectures. After normalizing for clock speed (2.2 GHz vs 1.8 GHz), the x86 platform still executes operators like \texttt{hash} with significantly fewer cycles, pointing to a superior microarchitectural efficiency that we will analyze next.

\begin{table}[htbp]
\centering
\caption{Performance Profiles of Operators on Arm and X86 Platforms (Base Cost is measured in CPU cycles for 64-byte packets)}
\label{tab:combined_profile}
\setlength{\tabcolsep}{5pt} 
\begin{tabular}{l l rrr}
\toprule
\textbf{Operator} & \textbf{Platform} & \textbf{Base Cost} & \textbf{Exponent (k)} & \textbf{$R^2$} \\
\midrule
\texttt{CRC}      & Arm & $\approx$823    & 1.3700 & 0.9976 \\
                  & x86 & $\approx$747    & 1.2699 & 0.9997 \\
\midrule
\texttt{Checksum} & Arm & $\approx$65     & 0.1632 & 0.9981 \\
                  & x86 & $\approx$27     & 0.1551 & 0.9995 \\
\midrule
\texttt{hash}     & Arm & $\approx$34     & 0.2606 & 0.9762 \\
                  & x86 & $\approx$9      & 0.1547 & 0.9988 \\
\midrule
\texttt{htons}    & Arm & $\approx$49     & 0.2067 & 0.9993 \\
                  & x86 & $\approx$1.5    & 0.0644 & 0.9634 \\
\midrule
\texttt{printf}   & Arm & $\approx$12,006 & 0.1130 & 0.9358 \\
                  & x86 & $\approx$29,129 & 0.2222 & 0.9561 \\
\midrule
\texttt{rte\_log}  & Arm & $\approx$108    & 0.2429 & 0.9327 \\
                  & x86 & $\approx$49     & 0.1509 & 0.9653 \\
\bottomrule
\end{tabular}
\end{table}

\begin{figure}[t!]
\centering
\includegraphics[width=\columnwidth]{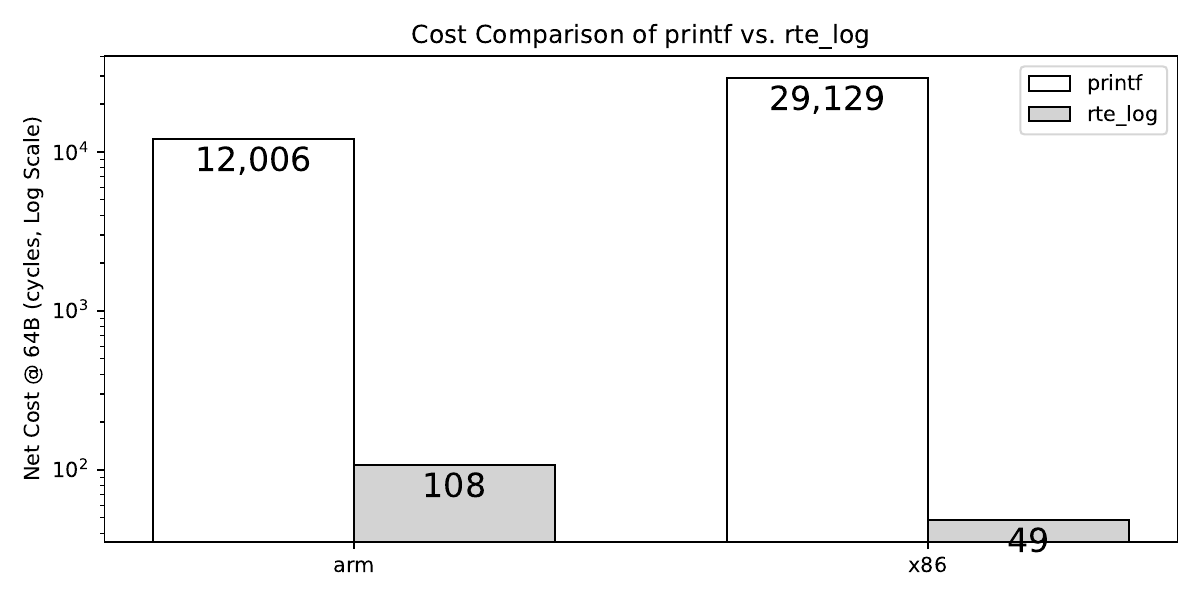}
\caption{Cost comparison between \texttt{printf} and \texttt{rte\_log} for 64-byte packets on both platforms. The Y-axis uses a log scale to show the vast performance gap.}
\label{fig:printf_comparison}
\end{figure}

\subsection{The Quadrant Shift Phenomenon}

When we project the data from both platforms onto the OPQ plot (Fig. 3), we can perform a quadrant-by-quadrant analysis based on the framework from Section II-C. This not only reveals performance shifts but also dictates specific optimization strategies.

\textbf{1) The Latent Traps Quadrant (High Base, k>1):}
The \texttt{CRC} operator is a classic example of a "Latent Trap," falling squarely into this quadrant on both architectures. Its high base cost and super-linear scaling make it a fundamental bottleneck. The OPQ analysis prescribes that for such operators, simple platform migration is insufficient; the only effective remedies are radical \textbf{algorithm replacement} or \textbf{hardware offloading}.

\textbf{2) The High Startup Cost Quadrant (High Base, k<1):}
On the Arm platform, \texttt{htons} and \texttt{Checksum} are classified into this quadrant as their base costs (~49 and ~65 cycles) exceed the global median. Their scaling is favorable (k<1), but the high fixed cost is the primary performance limiter. For applications bottlenecked by them, the framework suggests \textbf{batch processing} or vectorization to amortize this cost.

The \texttt{printf} operator represents an extreme case of this quadrant. Its base cost is orders of magnitude higher than any other operator (~12,006 cycles on Arm, ~29,129 on x86), making it a critical performance sink. Here, the prescribed strategy is not just amortization but complete \textbf{avoidance} in production data planes, in favor of lightweight logging alternatives like \texttt{rte\_log}.

\textbf{3) The Ideal Operators Quadrant (Low Base, k<1):}
This quadrant is the destination for optimization. On x86, \texttt{htons}, \texttt{hash}, and \texttt{Checksum} all shift into this ideal state. For example, \texttt{htons}'s base cost plummets to ~1.5 cycles, indicating profound microarchitectural optimization that solves the performance problem entirely. This shift from "High Startup Cost" to "Ideal" is a key finding, showing that a bottleneck on one platform can be a non-issue on another.

\textbf{4) The Emergent Bottlenecks Quadrant (Low Base, k>1):}
Although no operator in our dataset fell into this quadrant, it theoretically identifies Emergent Bottlenecks, which means operators with low base costs but super-linear scaling $(k>1)$ whose rapid performance degradation with larger packets would necessitate a hybrid algorithm strategy.


This quadrant-based analysis reinforces that performance is not portable. An application's bottleneck can shift from one quadrant's problem type to another's when migrating architectures, underscoring the necessity of a structured, architecture-specific analysis like OPQ.

\begin{figure}[t!]
\centering
\includegraphics[width=\columnwidth]{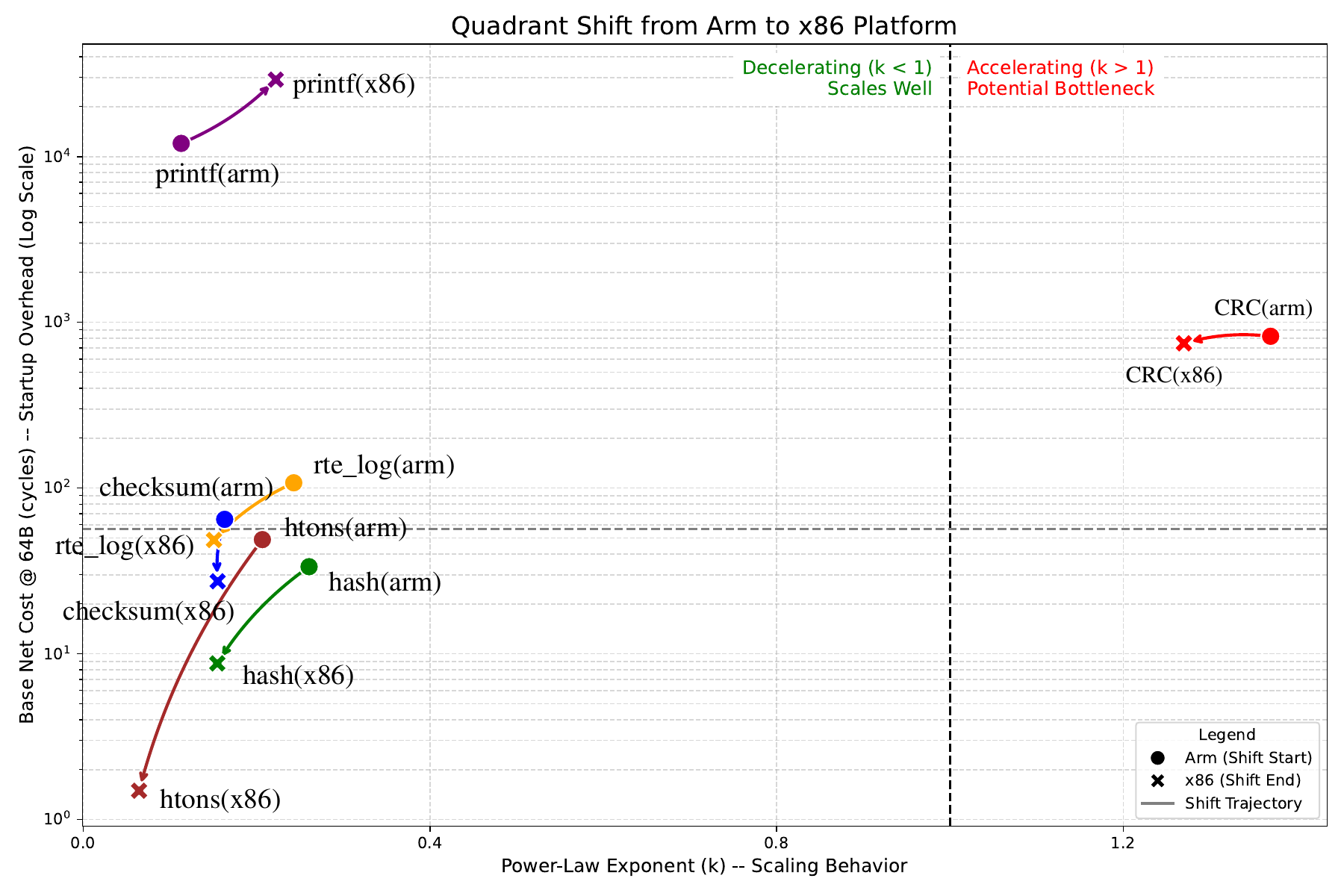}
\caption{The upgraded OPQ, plotting the power-law exponent $k$ against the Base Cost. Arrows indicate the performance shift of an operator from the Arm platform to the x86 platform, visually representing the Quadrant Shift phenomenon.}
\label{fig:quadrant_shift}
\end{figure}

\section{Conclusion}
This paper proposed and validated a diagnostic methodology to accurately assess operator costs in high-speed data planes. Our experiments on x86 and Arm platforms revealed complex non-linear scaling behaviors and the significant "Quadrant Shift" phenomenon, revising the traditional view of linear costs and demonstrating that performance optimizations are not architecturally portable. The OPQ framework was constructed to analyze these complex cost structures, and we quantified the immense optimization value in addressing pitfalls like \texttt{printf}. In summary, this work provides a more realistic diagnostic tool and theoretical basis for performance modeling and bottleneck prediction. 

Future work includes leveraging our non-linear cost data to train machine learning models (e.g., GNNs) for service chain performance prediction, and integrating the methodology into Data Plane as a Service (DPaaS) platforms for online, architecture-aware bottleneck analysis and scheduling.

\bibliographystyle{IEEEtran}
\bibliography{bibtex}

\end{document}